\documentclass[a4paper,fleqn,usenatbib]{mnras}
\usepackage{mathptmx}

\usepackage[T1]{fontenc}
\usepackage{ae,aecompl}

\usepackage{pdflscape}
\usepackage{longtable}
\usepackage{rotating}
\usepackage{ragged2e}

\usepackage{graphicx}	
\usepackage{amsmath}	
\usepackage{amssymb}	
\usepackage{booktabs}
\usepackage{float}
\usepackage{array}
\usepackage{wrapfig}
\usepackage{multirow}
\usepackage{tabu}
\usepackage{adjustbox}
\usepackage{tabularx}
\usepackage{booktabs,caption,fixltx2e}
\usepackage[flushleft]{threeparttable}
\usepackage{subcaption}
\usepackage{mathtools}

\usepackage[dvipsnames]{xcolor}
\hypersetup{colorlinks=true,linkcolor=Magenta,urlcolor=Plum,citecolor=DarkOrchid}

\newcommand{\HI}{\mbox {\sc H\thinspace{i}}}

\title[Peculiar Choir Galaxy ESO156-G029]{Group pre-processing versus cluster ram-pressure stripping: the case of ESO156-G029 }

\author[D\v{z}ud\v{z}ar et al.]{Robert D\v{z}ud\v{z}ar,$^{1}$\thanks{E-mail: rdzudzar@swin.edu.au; robertdzudzar@gmail.com}
Virginia Kilborn$^{1,2}$,
Chandrashekar Murugeshan$^{1,2}$,
\newauthor
Gerhardt Meurer$^{3}$,
Sarah M. Sweet$^{1,2}$,
Mary Putman$^{4}$
\\
$^{1}$Centre for Astrophysics and Supercomputing, Swinburne University of Technology, P.O. Box 218, Hawthorn, VIC 3122, Australia\\
$^2$ARC Centre of Excellence for All Sky Astrophysics in 3 Dimensions (ASTRO 3D)\\
$^3$International Centre for Radio Astronomy Research, ICRAR M468, 35 Stirling Highway, Crawley, WA 6009, Australia\\
$^4$Department of Astronomy, Columbia University, 550 West 120th Street, New York, 10027, USA }

\date{Accepted XXX. Received YYY; in original form ZZZ}

\pubyear{2019}

\begin{document}
\label{firstpage}
\pagerange{\pageref{firstpage}--\pageref{lastpage}}
\maketitle

\begin{abstract}
We report on observations of ESO156-G029, member of a galaxy group which is positioned at the virial radius of cluster Abell 3193. ESO156-G029 is located $\sim$ 1.4 Mpc in projected distance from the brightest cluster galaxy NGC1500. We show that ESO156-G029 has disturbed gas kinematics and a highly asymmetric neutral hydrogen (\HI) distribution, which are consequences of group pre-processing, and possibly of ram-pressure. Based on the current data we propose a scenario in which ESO156-G029 had a minor gas-rich merger in the past and now starts to experience ram-pressure. We infer that the galaxy will undergo rapid evolution once it gets closer to the cluster centre (less than 0.5 Mpc) where ram-pressure is strong enough to begin stripping the \HI\ from the galaxy.
\end{abstract}




\section{Introduction}

Galaxy clusters, the largest gravitationally bound structures, grow via mergers of smaller systems such as galaxies and galaxy groups \citep{Press1974, White1978}. Properties of galaxies in clusters are observed to change with respect to the environmental density (e.g. the morphology--density relation \citep{Dressler1980} and the star formation--density relation \citealt{Gomez2003}). Within galaxy clusters the typical velocities and intracluster medium (ICM) densities are sufficiently high that ram-pressure can strip the interstellar medium (ISM) from member galaxies, even leading to a complete removal of their cold gas as they move through the ICM \citep{GunnGott1972, Chung2009, Brown2017, Yoon2017}. 

Before galaxies fall into clusters they can go through transformations (e.g. change in the gas content and morphology), so-called `pre-processing', especially if those galaxies are part of groups \citep{McGee2009, Kilborn2009, Hess2013, Vijayaraghavan2013, Hou2014, Yoon2017}. Due to the small velocity dispersion in groups, galaxy mergers are frequently observed \citep{Hickson1997, Sengupta2017}. It has been shown that a galaxy's neutral hydrogen gas fraction can increase after a merger \citep{Ellison2018}, which can make the galaxy inefficient in forming stars due to an enhanced turbulence. A variety of physical processes can be efficient in galaxy groups that can shape the galaxy evolution (e.g. \citealt{Vulcani2018}). Thus, unraveling the impact of the group and the cluster on cluster infalling galaxies is an ongoing challenge \citep{Cortese2006, Hess2013, Hou2014, Hess2015}. 

In this work we investigate the physical processes that are shaping a spiral galaxy ESO156-G029. The location of ESO156-G029, falling into the cluster and part of a group, is excellent for gathering insights on galaxy pre-processing. 

This letter is structured as follows. Section \ref{data} describes the data. Section \ref{results} presents the galaxy group HIPASSJ0400-52, the cluster Abell 3193 and ESO156-G029 and the main results. We conclude in Section \ref{discussion}. Throughout this paper the assumed cosmology is H$_{0}$ = 70 km s$^{-1}$ Mpc$^{-1}$, $\Omega_{\text{m}}$ = 0.3, and $\Omega_{\Lambda}$ = 0.7, and we adopted \citep{Chabrier2003} initial mass function.

\begin{figure*}
\includegraphics[width=1\textwidth]{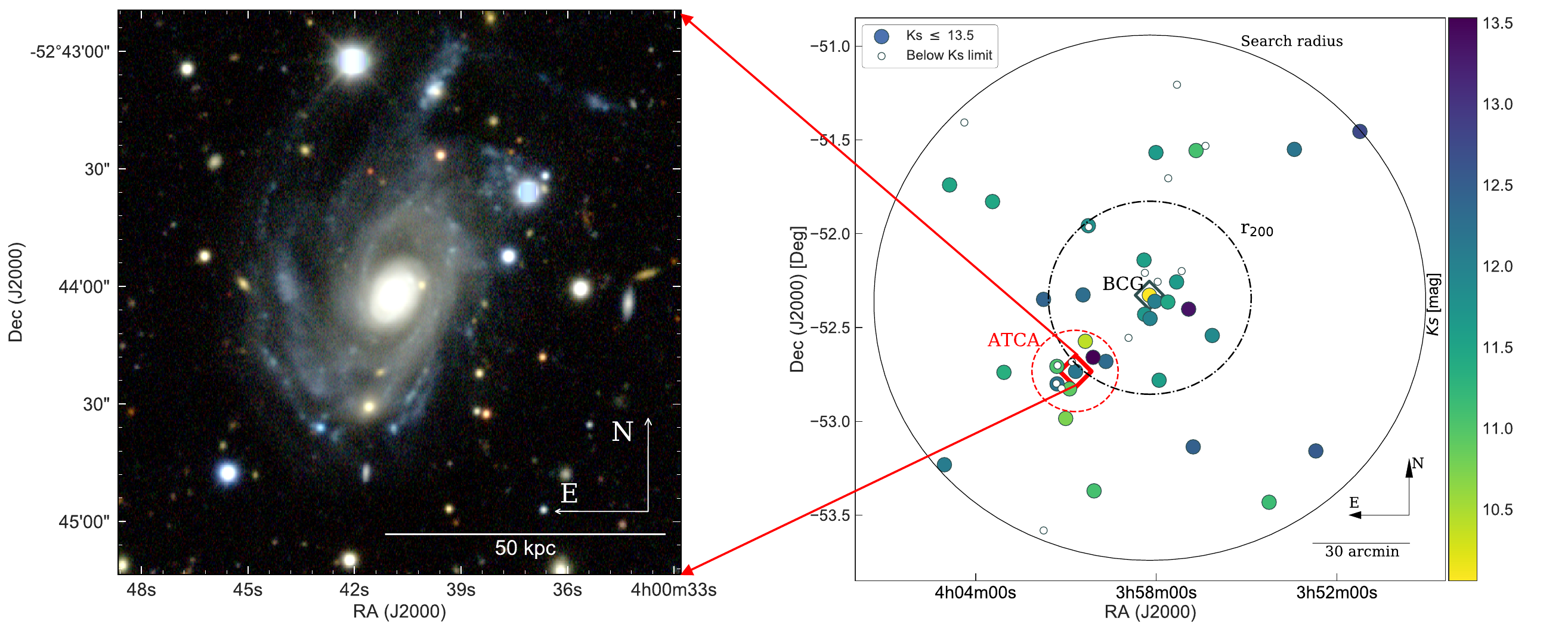}
\caption{\textbf{Left:} The DECam \textit{g}, \textit{r}, \textit{i} colour composite image of ESO156-G029. The blue star forming regions are mostly located in the disk outskirts. Faint stellar streams (\textit{r} $\sim$ 21.8--20.4 mag) are visible in the north-west of the galaxy. \textbf{Right:} 3$\times$3 degree region centered on Abell 3193. ESO156-G029 is marked with the red diamond. The cluster galaxies are shown as circles coloured based on their \textit{Ks} magnitude; shown galaxies are within the search radius (84 arcmin) from the cluster centre and within the velocity range of 9720--11690 km s$^{-1}$. } 
\label{fig:galaxy}
\end{figure*}

\section{Data}
\label{data}

ESO156-G029 was observed with the Australia Telescope Compact Array (ATCA) as part of the project C2440. The observation and data reduction of the ATCA data used are described in \citet{Dzudzar2019}. In summary, we used the standard reduction procedure in MIRIAD \citep{Sault1995}, create the \HI\ data cube with a channel width of 5 km s$^{-1}$, and a Brigg's robust weighting of 0.5 as used in \citet{Dzudzar2019}. We make use of the ATCA \HI\ intensity distribution of ESO156-G029, its \HI\ spectrum and its kinematic map.

The Two Micron All Sky Survey (2MASS) \textit{Ks}-band \citep{Skrutskie2006} background subtracted image is used to measure the \textit{Ks}-band luminosity which was converted into the stellar mass using \citet{Wen2013} relation. We use deep optical imaging (\textit{g}, \textit{r}, \textit{i}) from the first data release of the Dark Energy Survey (DES, \citealp{DES2018, Abbott2018}) obtained using Dark Energy Camera (DECam, \citealt{Flaugher2015}) on the CTIO Blanco 4-m telescope to trace faint features in ESO156-G029 (see Figure \ref{fig:galaxy}). 

\section{Results}
\label{results}

In this Section we present the properties of ESO156-G029 and its local and global environment in order to understand the physical processes that are acting on this galaxy. 
\subsection{The Choir group HIPASSJ0400-52}
\label{choir}
The Choir group HIPASSJ0400-52 (hereafter J0400-52) was identified in \citet{Sweet2013}, which showed general group properties obtained from the SINGG survey (Survey of Ionization in Neutral Gas Galaxies, \citealp{Meurer2006}). The adopted distance for J0400-52 group is 151 Mpc (z $=$ 0.035), and its coordinates RA: 04h00m40.82s; DEC: -52d44m02.72s \citep{Meurer2006, Sweet2013}. Using narrow-band emission imaging, nine emission-line galaxies were discovered in J0400-52: four spiral galaxies and five dwarf galaxies \citep{Sweet2013}. Spectroscopic follow-up observations yielded redshifts of these nine galaxies \citep{Sweet2014}, seven of them are confirmed J0400-52 members. Two galaxies have a higher recessional velocity ($\sim$ 1000 km s$^{-1}$) with respect to the mean group velocity. Thus, we exclude them as being group members, however, they remain as cluster members. Based on the recessional velocity and position, it appears that the J0400-52 group is infalling into the cluster Abell 3193 (hereafter A3193). Out of the seven known J0400-52 members, only ESO156-G029 was detected in \HI\ emission with the ATCA observations.   

\subsection{Abell 3193}
\label{abell}

The quoted properties of the cluster A3193 differ from study to study, thus we re-derive them. 
We search for galaxies within the radius of 1.4 deg\footnote{\url{https://www.cfa.harvard.edu/~dfabricant/huchra/clusters/table.html}} around the cluster centre, using NASA/IPAC Extragalactic Database (NED) and obtain a sample of 47 galaxies in the velocity range of 9720 and 11690 km s$^{-1}$. These limits were adopted based on the velocity dispersion of the galaxies around the cluster. We define a \textit{Ks} magnitude limited sample (33 galaxies brighter than the \textit{Ks} magnitude limit of 13.5 mag for extended sources; see Figure \ref{fig:galaxy}) to derive cluster properties. The mean cluster velocity is 10580$\pm$165 km s$^{-1}$ with a radial velocity dispersion of 540$\pm$42 km s$^{-1}$. We also find that the velocity of the brightest cluster galaxy (BCG), NGC1500 (Ks $\sim$ 10.07 mag), does not correspond to the mean cluster velocity, but instead is offset by $\sim$ 500 km s$^{-1}$. The kinematic offset of the BCG with respect to the cluster can indicate that the cluster is dynamically young and still in the process of relaxation \citep{Lauer2014, Wolfinger2016}.

From the radial velocity dispersion we derive the cluster radius and mass: r$_{200} =$ 1.3 Mpc and M$_{200} =$ 2.7$\times$10$^{14}$ M$_\odot$ (based on equations for r$_{200}$ and M$_{200}$ from \citealt{Poggianti2010}).  

From the ROSAT All-Sky Survey Faint Source Catalog \citep{Voges2000} A3193 has a weak X-ray emission (X-ray luminosity of 10$^{42}$ erg s$^{-1}$). In order to determine the extent of the cluster X-ray emission, we re-derive the X-ray counts from the ROSAT All-Sky Survey images. Using Mission Count Rate Simulator (WebPIMMS\footnote{\url{heasarc.gsfc.nasa.gov/cgi-bin/Tools/w3pimms/w3pimms.pl}}) we obtain the X-Ray luminosity of 1.6$\times$10$^{42}$ erg s$^{-1}$, which is in agreement with \citet{Voges2000}. The X-ray luminosity is localized within $\sim$ 3.5 arcmin around the cluster centre, coinciding with the position of NGC1500. 

\subsection{ESO156-G029}
\label{s1galaxy}

The integrated properties of ESO156-G029 were shown in \citet{Dzudzar2019} - galaxy with ID: HIPASSJ0400-52:S1. We found that ESO156-G029 is an \HI-rich galaxy with a large gas-mass fraction and a low specific star formation efficiency. We analytically determined that ESO156-G029 has a high specific angular momentum, which may hinder the ISM from forming stars, and thus causing low star formation efficiency in the galaxy; in a similar manner to \HI\ extreme galaxies \citep{Lutz2017, Lutz2018}. We use the derived star formation rate of ESO156-G029 from H$\alpha$ and the Wide Field Imaging Survey Explorer (WISE; using 12$\mu$m, \citet{Jarrett2013}) and find that ESO156-G029 lies on the star forming main sequence when comparing to the sSFR main sequence of \citet{Catinella2018}. 

\begin{figure}
\textbf{(a)}
\centering
\includegraphics[width=.71\columnwidth]{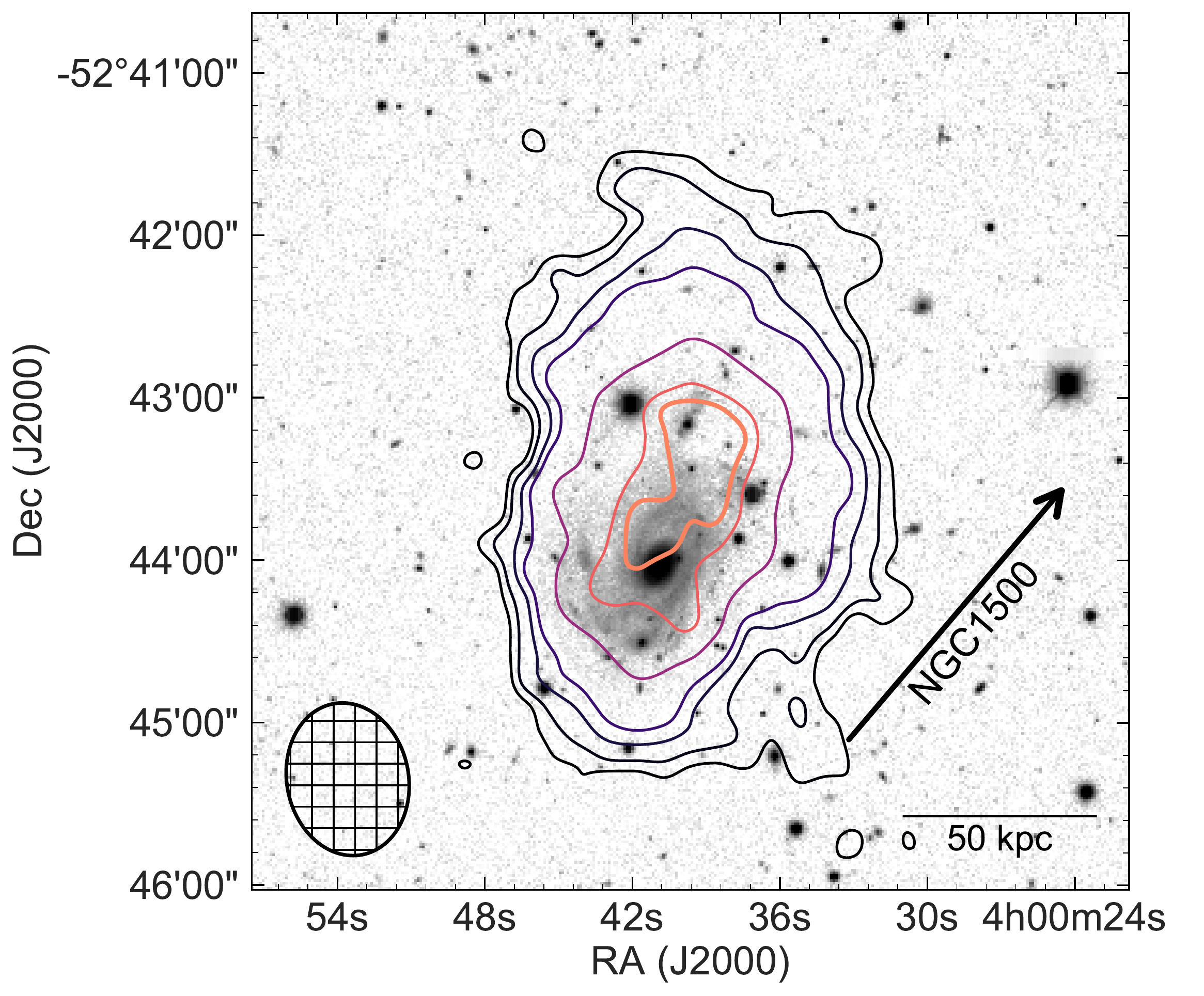}\\
\textbf{(b)}
\includegraphics[width=.71\columnwidth]{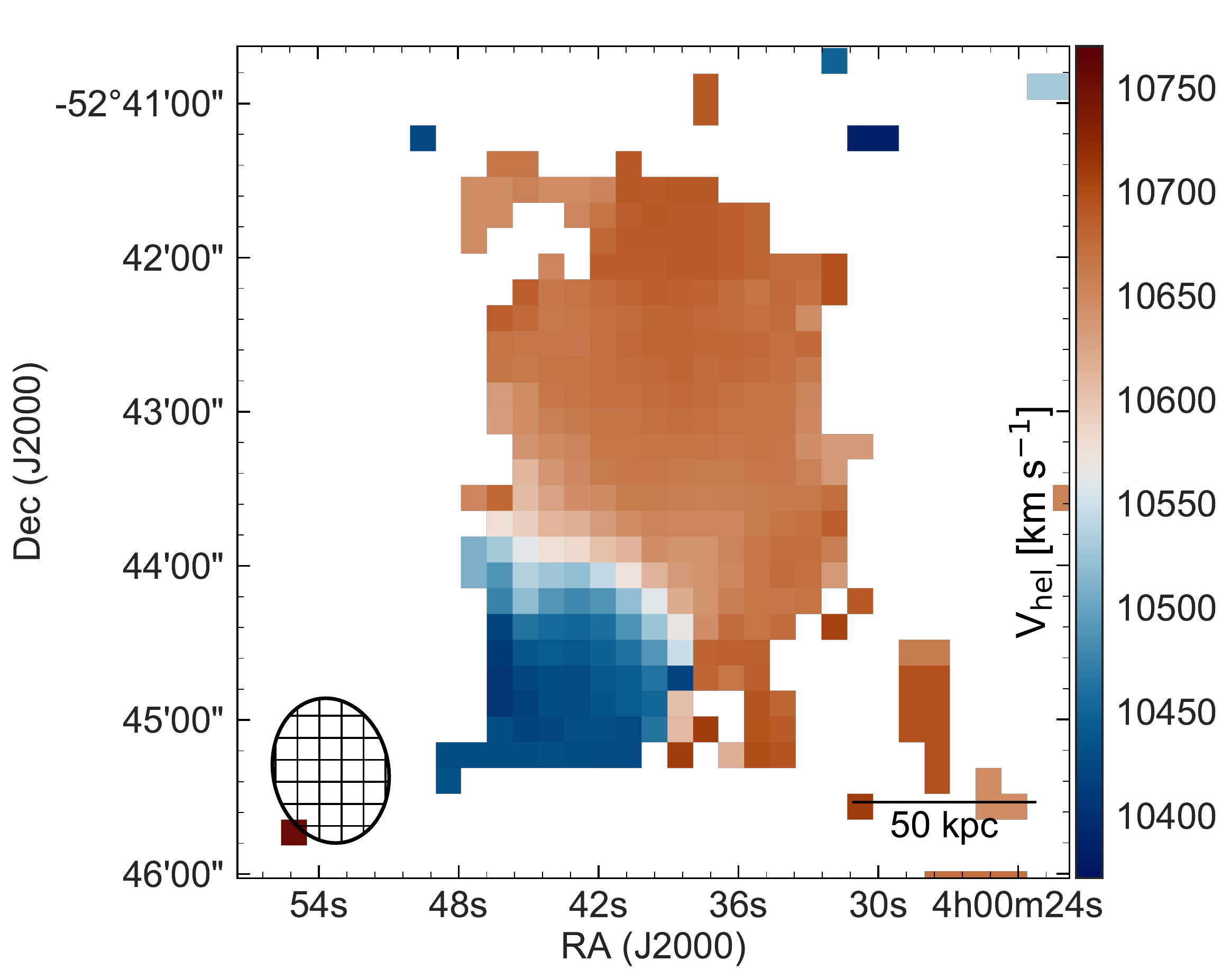}\\
\textbf{(c)}
\includegraphics[width=.58\columnwidth]{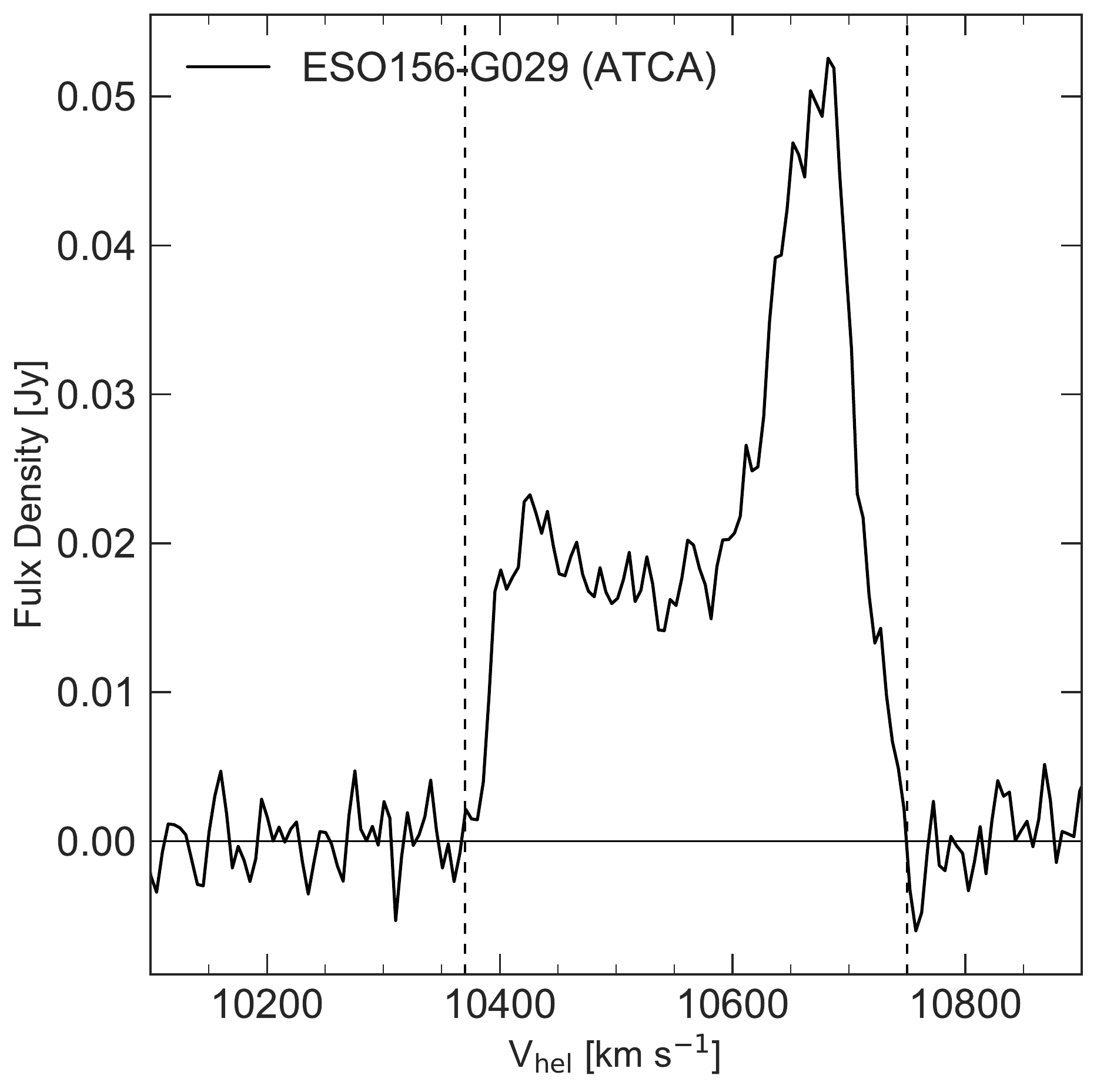}\\
\textbf{(d)}
\includegraphics[width=.67\columnwidth]{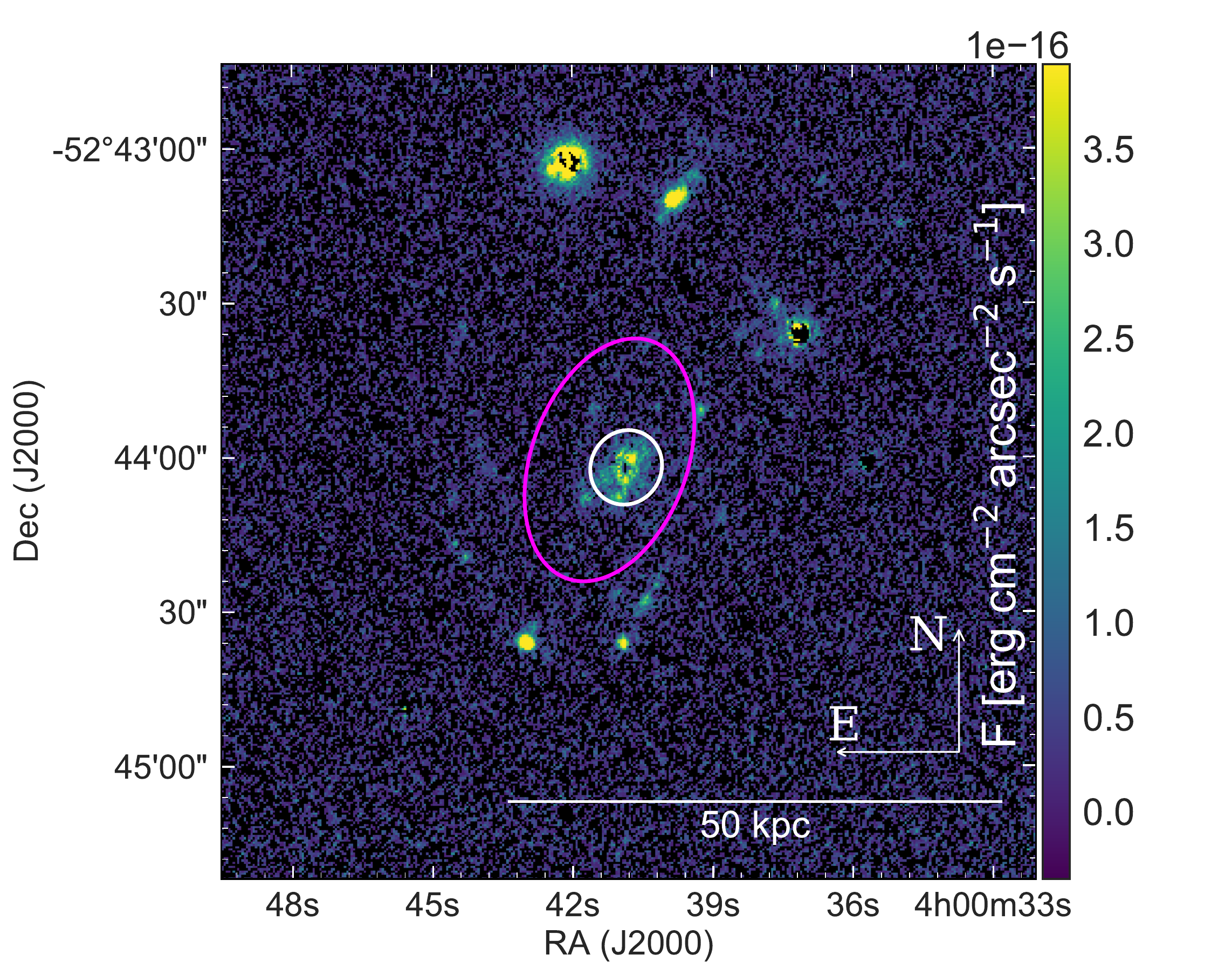}
\caption{{\textbf{(a)}} The \HI\ emission of ESO156-G029 is shown by the contours overlaid on the DECam {\textit{g}} band image. The lowest shown \HI\ column density corresponds to 3$\times$10$^{19}$ cm$^{-2}$. The other shown contours have column densities of: 12, 18, 36, 54 and 60$\times$10$^{19}$ cm$^{-2}$. \textbf{(b)} The velocity map of ESO156-G029. The synthesized beam (45$\arcsec$ $\times$ 57$\arcsec$) is shown in the bottom left corner of panel \textit{a} and \textit{b}. \textbf{(c)}
The \HI\ spectrum of ESO156-G029, extracted from the ATCA data cube. The black dashed vertical lines show the velocity range over which the \HI\ spectrum was integrated. \textbf{(d)} ESO156-G026 in H$\alpha$ imaging (SINGG; \citealt{Meurer2006}). The colourbar shows the surface brightness. The white circle marks galaxy's bulge and the magenta ellipse denotes region between inner and outer disc (see Section \ref{sec:interaction}).} 
\label{fig:HI_prop_gal}
\end{figure}

ESO156-G029, whilst being a part of the group J0400-52, it is also a cluster infall galaxy, thus it is an excellent case study into the interplay between group and cluster environment. We discuss the physical processes that are responsible for the peculiar features of ESO156-G029: the highly asymmetric \HI\ profile, disturbed \HI\ kinematics and star-forming stellar streams (seen in optical and H$\alpha$; Figure \ref{fig:galaxy} and \ref{fig:HI_prop_gal}d).

\begin{table}
\centering
\begin{threeparttable}
\caption{Properties of ESO156-G029}
\label{table:s1_properties}
\begin{tabular}{p{4.5cm} p{2.5cm}}
\toprule \hline 
Right Ascension (optical)\tnote{a,b}   &   04h00m40.82s       \\  
Declination (optical)\tnote{a,b}       &   -52d44m02.71s       \\
V$_{\textrm{hel, \HI}}$ (km s$^{-1}$) &   10570      \\
V$_{\textrm{hel, H$\alpha$}}$ (km s$^{-1}$)\tnote{a} &   10424      \\
D (Mpc)\tnote{a,b}  &     151       \\
z \tnote{a} & 0.0348 \\
D$_{\textrm{proj, NGC1500}}$ (Mpc)    &    1.4   \\
SFR$_{\textrm{H}\alpha}$ (M$_{\odot}$ yr$^{-1}$)\tnote{b}     &    0.5$\pm$0.1\tnote{*}   \\
SFR$_{\textrm{W12}}$ (M$_{\odot}$ yr$^{-1}$)       &   2.34$\pm$0.82    \\
V$_{\textrm{flat}}$ (km s$^{-1}$) & 142.2$\pm$7.4 \\
$\sigma_{\textrm{\HI}}$ (km s$^{-1}$) & 22$\pm$8\\
i (deg) & 60.3$\pm$1.4 \\
PA (deg) & 341$\pm$3  \\
f$_{\textrm{atm}}$ & 0.49$\pm$0.04 \\
log j (M$_{\odot}$ km s$^{-1}$ kpc) & 3.51$\pm$0.08 \\
q & 0.16$\pm$0.05 \\
log M$_{\HI}$ (M$_\odot$)\tnote{$\dagger$} & 10.6$\pm$0.1 \\
log M$_{\star}$($_{\textrm{Ks}}$) (M$_\odot$)          &   10.7$\pm$0.1        \\ \bottomrule \hline
\end{tabular}
\begin{tablenotes}
\item a - \citep{Sweet2014}; b -  \citep{Dzudzar2019}. All other properties are derived in this work.\\ $\dagger$ The quoted \HI\ mass is unclipped while in \citet{Dzudzar2019} we used 3$\sigma$ clipping. \item * \citep{Dzudzar2019} quotes SFR error of 0.4 in the Table, which is an actual error, it should be 0.1.

\end{tablenotes}
\end{threeparttable}
\end{table}

\subsubsection{The \HI\ distribution and kinematics of ESO156-G029}

We show the \HI\ distribution of ESO156-G029 overlaid on the DECam \textit{g} band image (see Figure \ref{fig:HI_prop_gal}a). The \HI\ distribution has an offset of $\sim$ 20 arcsec from the stellar centre of the galaxy. The \HI\ contours in the south-east part of the galaxy are compressed with respect to those on the north-west. In panel \textit{b} of Figure \ref{fig:HI_prop_gal}, the velocity map of ESO156-G029 shows that the gas in the galaxy is being skewed towards higher velocities, which is possibly caused by the ram-pressure of the cluster. We also find that the rotation curves in the approaching and receding sides of the galaxy are very different, with a difference in maximum velocities of the order of $\sim$ 40 km s$^{-1}$. The \HI\ asymmetry is also evident in the emission line spectrum in panel \textit{c} of Figure \ref{fig:HI_prop_gal}. The peak of the integrated \HI\ spectrum for the receding side of the galaxy is $\sim$ 44 per cent higher than the peak for the approaching side. Following \citet{Espada2011} we derive the integrated density flux ratio parameter ($A_{flux\;ratio}$, as a measure of asymmetry; the higher the value than 1.0, the higher the asymmetry) to be $\sim$ 1.7, which places this galaxy into the class of those with strongly asymmetric \HI\ profiles (only four out of 318 galaxies from the \citet{Espada2011} sample have similarly high asymmetry values). Such an asymmetry could be a result of a past merger (e.g. \citealp{Holwerda2011, Scott2018, Bok2019}).

\subsubsection{Modeling of ESO156-G029}

We use the code 3DBarolo (3D-Based Analysis of Rotating Object via Line Observations, \citealt{DiTeodoro2015}) to fit the \HI\ emission-line data-cube of ESO156-G029. Kinematic and geometrical parameters (velocity dispersion ($\sigma_{\textrm{\HI}}$), inclination (i), position angle (PA)) obtained from the modelling are given in Table \ref{table:s1_properties}. We obtain the maximum rotational velocity of the galaxy (velocity where the rotation curve is flat, V$_{\textrm{flat}}$). 

Using data from the \HI\ modeling and the \textit{Ks}-band (background subtracted) galaxy image, we derive the total baryonic specific angular momentum (j) and stability parameter (q) of the galaxy utilizing the robust method described in \citet{Murugeshan2019}. We determine that the global stability parameter (q) is in overall agreement with the one determined analytically by \citet{Dzudzar2019}. \citet{Obreschkow2016} observe a tight positive correlation between the atomic gas-mass fraction and the global stability parameter q (with its ability to regulate star formation efficiency) which consequently determines the atomic gas fraction. Based on the q value, we can conclude that ESO156-G029 has a high angular momentum and thus low star formation efficiency. The properties of ESO156-G029 are similar to the galaxies studied by \citet{Gereb2018} and \HI\ eXtreme galaxies (HIX, \citealt{Lutz2017, Lutz2018}), although ESO156-G029 shows signs of a disturbance. 
In the following sections we explore whether ram-pressure and galaxy interaction may have had an effect on ESO156-G029.

\begin{figure}
\centering
\includegraphics[width=0.85\columnwidth]{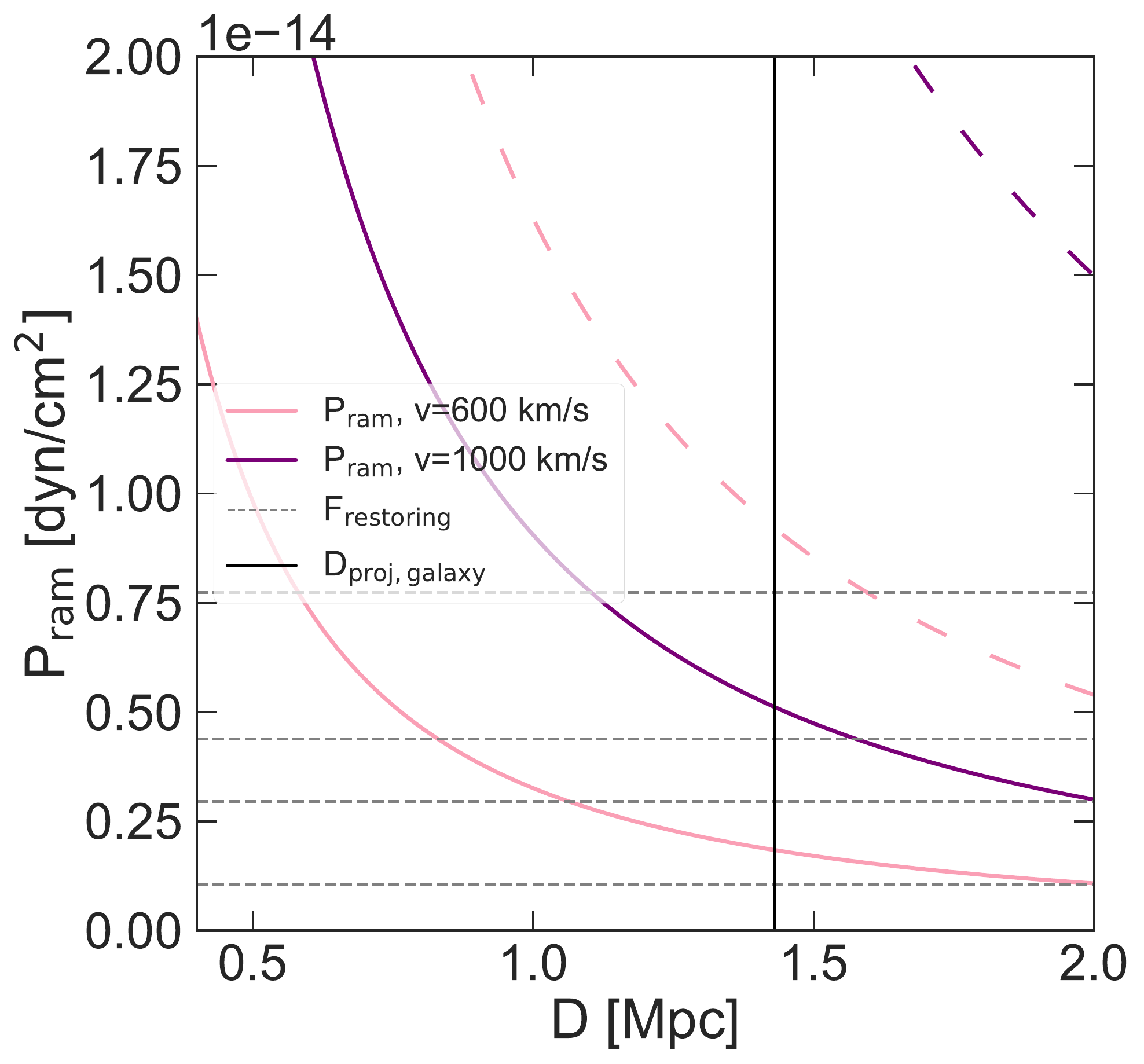}
\caption{Comparison of the ram-pressure and gravitational restoring force for ESO156-G029. The vertical line is the galaxy's projected distance from the cluster centre. The curved lines correspond to the ram-pressure computed for two different velocities: 600 and 1000 km s$^{-1}$; the dashed curved lines assume ICM central density of 5$\times$10$^{-4}$ cm$^{-3}$, while the solid curved lines assume ICM central density of 10$^{-4}$ cm$^{-1}$. The horizontal dashed lines correspond to the galaxy's restoring force, computed for the \HI\ surface densities of: 0.2, 0.4, 0.45 and 0.65 M$_{\odot}$ pc$^{-2}$ (respectively, from to bottom to the top).
}
\label{fig:Pram}
\end{figure}

\subsection{Ram-pressure stripping}

As galaxies move through the intra cluster medium (ICM) they can be affected by ram-pressure ($P_{\textrm{ram}}$) of the ICM: $P_{\textrm{ram}}=\rho_{\textrm{ICM}} v^2_{g} $; where $\rho_{\textrm{ICM}}$ is the ICM density and $v$ is the galaxy velocity with respect to the ICM \citep{GunnGott1972}. The extreme effect of ram-pressure can cause stripping of the gas content off galaxies if the ram-pressure exceeds the restoring force of the galaxy, i.e. P$_{\textrm{ram}}$ \textgreater  2$\pi$G$\Sigma_{\textrm{tot}}\Sigma_{\textrm{gas}}$. We use rough approximations to estimate whether ram-pressure is effective (e.g. \citealt{Bravo2009}), utilizing only the surface mass densities of the \HI\ ($\Sigma_{\textrm{gas}}$) and stars ($\Sigma_{\textrm{tot}} = \Sigma_{\textrm{gas}} + \Sigma_{\textrm{stars}}$). We use the hydrostatic-isothermal $\beta$-model of \citet{Cavaliere1976} to find the ICM density distribution: $\rho (r) = \rho_{0}\left [ 1 + (r/r_{c})^{2} \right ]^{-3\beta/2}$; using $\beta$=0.5, cluster core radius r$_{c}\sim$ 38 kpc (at 50 per cent of the peak X-ray surface brightness) and central density 10$^{-4}$ \textless $\rho_{0}$ \textless 5$\times$10$^{-4}$
cm$^{-3}$. The ram-pressure parameter P$_{\textrm{ram}}$ is derived for two extreme ranges of the central ICM density, as well as two values of the galaxy velocity. We compare the strength of the ram-pressure stripping with the restoring force of the galaxy (see Figure \ref{fig:Pram}), at the galaxy's projected distance from the cluster centre $\sim$ 1.4 Mpc. The four horizontal dashed lines in Figure \ref{fig:Pram} represent the restoring force of the galaxy \HI\ `edge' (computed for the four \HI\ surface densities: 0.2, 0.4, 0.45 and 0.65 M$_{\odot}$ pc$^{-2}$). For a reasonable approximation of the central ICM density and galaxy velocity with respect to the cluster centre (cluster velocity dispersion) we obtain that the restoring force of the galaxy \HI\ `edge' is similar to ram-pressure force. This leads us to the conclusion that if the galaxy is under the influence of ram-pressure it has to be an early phase ram-pressure since the galaxy is not gas-poor. Assuming that the galaxy is going towards cluster centre, based on the current projected position of ESO156-G029 and direction of the \HI\ asymmetry, the galaxy would have to be infalling onto the cluster from behind.

We find that ram-pressure will be much more stronger if galaxy approaches closer than 0.5 Mpc to the cluster centre, where we expect significant \HI\ stripping. An example of advanced stage of ram-pressure is galaxy JO206, which exhibits low \HI-gas fraction and enhanced star formation due to ram-pressure stripping \citep{Ramatsoku2019}. The advanced ram-pressure phase of ESO156-G029 would resemble the case from \citep{Ramatsoku2019} or even to the more extreme case, such as described in \citet{Fumagalli2014}, forming extended gaseous tail. 

\subsection{Galaxy--galaxy interaction}
\label{sec:interaction}
Galaxy interactions can lead to asymmetries in their \HI\ distribution and kinematics (e.g. \citealt{Holwerda2011}). It is also shown that galaxies which had a merger can exhibit an \HI-gas fraction enhancement \citep{Ellison2018}.

We find indications that ESO156-G029 is in an advanced (or post) merger stage. Firstly, the galaxy has a higher \HI-gas fraction when compared to galaxies of similar stellar mass \citep{Dzudzar2019}. Secondly, the majority of the star forming regions lie in the outer disc of the galaxy (see panel \textit{d} of Figure \ref{fig:HI_prop_gal}). The sum of H$\alpha$ flux in the outer disc regions is around two times greater than that from the inner regions (excluding the bulge). We also find the faint stellar streams in the north-west part of the galaxy (see Figure \ref{fig:galaxy}) which may be remnants of the shredded dwarf that has merged into the system. Since ram-pressure is very low, it is more likely that such optical morphology is a result of a past merger event. Moreover, the increased \HI\ velocity dispersion (see $\sigma_{\HI}$ in Table \ref{table:s1_properties}) is indicative of a turbulent medium, possibly due to a merger \citep{Elmegreen1995, Tamburro2009}. Also, galaxies exhibiting large differences in their (approaching versus receding) rotation curves are likely to have had a tidal encounter \citep{Eymeren2011}. To quantify the degree of tidal influence \citet{Eymeren2011} use the kinematic lopsidedness parameter (Eq. 1 in their paper). We apply this method to ESO156-G029 and find a high value of $\sim$ 0.13 for the kinematic lopsidedness parameter (normal values are expected around and below the mean of 0.056 \citep{Eymeren2011}), hinting to the fact that this galaxy may have had an encounter in its recent past.

We hence conclude that properties such as the optical morphology, \HI\ content, velocity dispersion and kinematic lopsidedness are consistent with galaxy-galaxy interaction rather than ram-pressure stripping.

\section{Discussion and conclusions}
\label{discussion}

We have presented observations of the spiral galaxy ESO156-G029 which reveal an asymmetric neutral hydrogen (\HI) gas distribution and disturbed gas kinematics that we have discovered in the J0400-52 galaxy group (HIPASSJ0400-52:S1, \citealt{Sweet2013, Dzudzar2019}). 

We argue that galaxy pre-processing, through an ongoing minor merger within the J0400-52 group is more likely to have caused the rare distinct features of this system and that (if present) the effect of ram-pressure is very low. We show that the galaxy, assuming it is moving towards cluster centre, will suffer significant effects of ram-pressure stripping when it gets closer than 0.5 Mpc from the cluster centre. Assuming that the galaxy is moving with a velocity of 600 km s$^{-1}$ towards cluster centre, it will take \textgreater 0.6 Myr before ram-pressure becomes capable of stripping the ISM of ESO156-G029. 

The majority of galaxies that reside within group and cluster environments are built-up from infalling galaxy groups and individual galaxies. Thus identifying more galaxies like ESO156-G029 which appear to be distorted as they enter the cluster environment will be crucial for allowing us to dissect which processes are responsible for their transformation.

\section{Acknowledgements}

We thank the anonymous referee for their comments and suggestions which improved this paper.\\
\indent Robert D\v{z}ud\v{z}ar is supported by the Swinburne University Postgraduate Award (SUPRA).\\
\indent Robert would like to thank Manodeep Sinha, Michelle Cluver, Tom Jarrett (for providing WISE photometry) Uro\v{s} Mestri\'{c}, Katharina Lutz, Vaishali Parkash, Jacob Seiler, Enrico di Teodoro and MicAstros for useful comments and discussions.\\
\indent Parts of this research were supported by the Australian Research Council Centre of Excellence for All Sky Astrophysics in 3 Dimensions (ASTRO 3D), through project number CE170100013.\\
\indent The Australia Telescope Compact Array is part of the Australia Telescope National Facility which is funded by the Australian Government for operation as a National Facility managed by CSIRO.\\
\indent This research has made use of the NASA/IPAC Ex- tragalactic Database (NED), which is operated by the Jet Propulsion Laboratory, California Institute of Technology, under contract with the National Aeronautics and Space Administration.\\
\indent This publication makes use of data products from the Two Micron All Sky Survey, which is a joint project of the University of Massachusetts and the Infrared Processing and Analysis Center/California Institute of Technology, funded by the National Aeronautics and Space Administration and the National Science Foundation.\\
\indent This project used public archival data from the Dark Energy Survey (DES). Funding for the DES Projects has been provided by the U.S. Department of Energy, the U.S. National Science Foundation, the Ministry of Science and Education of Spain, the Science and Technology FacilitiesCouncil of the United Kingdom, the Higher Education Funding Council for England, the National Center for Supercomputing Applications at the University of Illinois at Urbana-Champaign, the Kavli Institute of Cosmological Physics at the University of Chicago, the Center for Cosmology and Astro-Particle Physics at the Ohio State University, the Mitchell Institute for Fundamental Physics and Astronomy at Texas A\&M University, Financiadora de Estudos e Projetos, Funda{\c c}{\~a}o Carlos Chagas Filho de Amparo {\`a} Pesquisa do Estado do Rio de Janeiro, Conselho Nacional de Desenvolvimento Cient{\'i}fico e Tecnol{\'o}gico and the Minist{\'e}rio da Ci{\^e}ncia, Tecnologia e Inova{\c c}{\~a}o, the Deutsche Forschungsgemeinschaft, and the Collaborating Institutions in the Dark Energy Survey.
The Collaborating Institutions are Argonne National Laboratory, the University of California at Santa Cruz, the University of Cambridge, Centro de Investigaciones Energ{\'e}ticas, Medioambientales y Tecnol{\'o}gicas-Madrid, the University of Chicago, University College London, the DES-Brazil Consortium, the University of Edinburgh, the Eidgen{\"o}ssische Technische Hochschule (ETH) Z{\"u}rich,  Fermi National Accelerator Laboratory, the University of Illinois at Urbana-Champaign, the Institut de Ci{\`e}ncies de l'Espai (IEEC/CSIC), the Institut de F{\'i}sica d'Altes Energies, Lawrence Berkeley National Laboratory, the Ludwig-Maximilians Universit{\"a}t M{\"u}nchen and the associated Excellence Cluster Universe, the University of Michigan, the National Optical Astronomy Observatory, the University of Nottingham, The Ohio State University, the OzDES Membership Consortium, the University of Pennsylvania, the University of Portsmouth, SLAC National Accelerator Laboratory, Stanford University, the University of Sussex, and Texas A\&M University.
Based in part on observations at Cerro Tololo Inter-American Observatory, National Optical Astronomy Observatory, which is operated by the Association of Universities for Research in Astronomy (AURA) under a cooperative agreement with the National Science Foundation.\\
\indent This research has made use of \texttt{python} \url{https://www.python.org} and python packages: \texttt{astropy} \citep{Astropy2013, Astropy2018}, \texttt{matplotlib} \url{http://matplotlib.org/} \citep{Hunter2007}, \texttt{APLpy} \url{https://aplpy.github.io/} and \texttt{NumPy} \url{http://www.numpy.org/} \citep{VanDerWalt2011}. Figure \ref{fig:HI_prop_gal}b uses scientific colour map `vik' \citep{Crameri}. 

\bibliographystyle{mnras}
\bibliography{ESO156-G029} 

\end{document}